\documentclass[twocolumn,showpacs,preprintnumbers,superscriptaddress,a
msmath,amssymb]{revtex4}

\usepackage{graphicx}
\usepackage{dcolumn}
\usepackage{bm}
\usepackage{amsmath}

\newcommand{\be}{\begin{equation}}
\newcommand{\ee}{\end{equation}}
\newcommand{\ba}{\begin{eqnarray}}
\newcommand{\ea}{\end{eqnarray}}

\begin{document}

\preprint{APS preprint}

\title{On Statistical Methods of Parameter Estimation \\for Deterministically
Chaotic Time-Series}

\author{V.F. Pisarenko}
\affiliation{International Institute of Earthquake Prediction Theory and
Mathematical Geophysics, Russian Ac. Sci. Warshavskoye sh., 79, kor. 2,
Moscow 113556, Russia}

\author{D. Sornette}
\affiliation{Institute of Geophysics and Planetary Physics,
University of California, Los Angeles, CA 90095}
\affiliation{Department of Earth and Space Sciences, University of
California, Los Angeles, CA 90095\label{ess}}
\affiliation{Laboratoire de Physique de la Mati\`ere Condens\'ee,
CNRS UMR 6622 and Universit\'e de Nice-Sophia Antipolis, 06108
Nice Cedex 2, France}
\email{sornette@moho.ess.ucla.edu}

\date{\today}

\begin{abstract}
We discuss the possibility of applying some standard statistical
methods (the least square method, the maximum likelihood method, 
the method of statistical moments for
estimation of parameters) to deterministically chaotic
low-dimensional dynamic system (the logistic map) containing an
observational noise. A ``pure'' Maximum Likelihood (ML) method is
suggested to estimate the structural parameter of the logistic map along
with the initial value $x_1$ considered as an additional unknown parameter.
Comparisons with previously proposed techniques on simulated numerical
examples give favorable results (at least, for the investigated combinations of
sample size $N$ and noise level). Besides, unlike some suggested
techniques, our method does not require the a priori knowledge of the noise
variance. We also clarify the nature of the inherent difficulties in
the statistical analysis of deterministically chaotic time series and the
status of previously proposed Bayesian approaches. We note the
trade-off between the need of using a large number of data
points in the ML analysis to decrease the bias (to guarantee consistency of
the estimation) and the unstable nature of dynamical trajectories with
exponentially fast loss of memory of the initial condition. The method
of statistical moments for the estimation of the parameter of the logistic map
is discussed. This method seems to be the unique method whose
consistency for deterministically chaotic time series is proved so far
theoretically (not only numerically).
\end{abstract}


\maketitle

The problem of characterizing and 
quantifying a noisy nonlinear dynamical chaotic system from 
a finite realization of 
a time series of measurements is full of difficulties. The first one
is that one rarely has the luxury of knowing the underlying dynamics, i.e., 
one does not in general know the underlying equations of evolution. 
Techniques to reconstruct a parametric representation of the time series
then may lead to so-called model errors. 

Even in the rare situations where one can ascertain that the 
measurements correspond to a known set of equations with additive noise,
the chaotic nature of the dynamics makes the estimation of the 
model parameters from time series surprisingly difficult. This is true
even for low-dimensional systems, another even rarer instance in 
naturally occurring time series.

Here, we revisit the problem proposed by McSharry and Smith
\cite{McSharrySmith}, who introduced an improved method over standard
least-square fits to estimate
the structural parameter of a low-dimensional deterministically chaotic system
(the logistic map). We discuss the caveats underlying this problem, propose
a ``pure'' Maximum Likelihood method that we compare with previously proposed
methods. Our conclusion stresses the inherent difficulties in formulating
a bona fide statistical theory of structural parameter estimations for
noisy deterministic chaos.

\section{Definition and nature of the problem}

Let us consider the supposedly simple problem considered by McSharry and Smith
\cite{McSharrySmith}, in which one measures the sample
$s_1, ..., s_N$ with
\be
s_i = x_i + \eta_i
\label{mgmwlw}
\ee
where the underlying dynamical one-dimensional discrete recurrence equation 
\be
x_{i+1} = F(x_i, a) \equiv 1 - a x_i^2~
\label{jjgled}
\ee
is known and the $\eta_i$'s are Gaussian $N(0,\epsilon)$ iid random variables 
with zero mean and standard deviation $\epsilon$. The problem is
to determine the model parameter $a$ from the measurements $s_1, ..., s_N$,
knowing that (\ref{jjgled}) is the true dynamics.

At first sight, this problem looks like a statistical estimation of 
an unknown structural parameter, given observational data.
However, strictly speaking, this problem cannot be (even formally) 
refered to as a bona fide statistical problem in which the
maximum likelihood (ML) method can be proved to be asymptotically optimal
or even consistent.
Indeed, the Likelihood Function $L(a,x_1| s_1,...,s_N)$ reads
\be
\ln L(a,x_1| s_1,...,s_N) \propto -N \ln(\epsilon) - 
{1 \over 2 \epsilon^2} \sum_i \left( s_i-F^{(i)}(x_1, a) \right)^2~,
\label{iwqsm}
\ee
where $F^{(i)}(x_1, a) $ is the $i$-th iteration of the logistic map 
(\ref{jjgled}) with parameter $a$ and initial value $x_1$. 
The key point of difficulty is that $F^{(i)}(x_1, a)$ is a 
{\it non-stationary} function
(despite the fact that the dynamical system (\ref{jjgled}) has an invariant
measure $\mu(x)$). Standard statistical ML methods are applicable
either to functions not depending on $i$, or depending on $i$ in a
periodic manner. For non-stationary and
non-periodic dependence of the function on $i$, no statistical theorem on
optimal properties of MLE is a priori applicable. Then, numerical simulations of
examples are not enough and should be complemented with proofs 
of results stating what known mathematical statistics properties
of ML or of Bayesian methods continue to apply to (\ref{iwqsm}). 
A first taste of the difficulty of the problem is given
by an analysis of the behavior of the ``one-step least-square (LS)
estimation'' and of the ``total least-square'' method, given in Appendix A.
Appendix A shows that least-square methods are biased and should be corrected
before comparing these to other methods, as done in \cite{McSharrySmith}.
In particular, Appendix A shows that it was a priori unfair or inappropriate
to compare any estimate obtained with a given 
method (such as the one advocated by McSharry and Smith
\cite{McSharrySmith}) to uncorrected ML-estimates due to the non-stationarity
of the function; the appropriate corrections can be obtained from the standard
statistical theory of confluence analysis \cite{FrischR,Geary,KendallStuart}.

\section{A ``pure'' Maximum Likelihood approach in terms of $(a, x_1)$ \label{nmgl}}

Putting aside the question of a rigorous demonstration of the consistency
and asymptotic optimality of the MLE method, let us come back to
expression (\ref{iwqsm}), which is the straightforward translation of the
iid Gaussian $N(0,\epsilon)$ properties of the random variables $\eta_i$'s.
It suggests that the problem of estimating the structural parameter $a$
cannot actually be separated from estimating simultaneously the
initial value $x_1$.

The MLE of $(a, x_1)$ amounts in this
case to the minimization of the sum:
\be
\sum_i \left( s_i-F^{(i)}(x_1, a) \right)^2~,
\label{mglle}
\ee
which looks superficially as a standard non-linear least-square sum.
There is however one very important distinction, as we already pointed out above:
the non-linear function depends
on the index $i$ whereas, in the standard least-square method, one has a sum of the type
\be
\sum_i \left( s_i-F(x_i, a) \right)^2~,
\label{mgllaaae}
\ee
where the $x_1, ..., x_N$ are assumed to be known.

For the parameters $a$ for which the logistic map exhibits
the phenomenon of sensitivity upon the initial condition, the direct
minimization of (\ref{mglle}) is not feasible directly. Indeed,
if we disturb $x_1$ by a small number $\delta$, then 
the $i$th iterations $F^{(i)}(x_1, a)$ and $F^{(i)}(x_1+\delta, a)$  diverge
asymptotically exponentially fast with $i$: for instance, 
with an accuracy $\delta = 10^{-15}$ and for $a=1.85$, the difference 
$F^{(i)}(x_1, a) - F^{(i)}(x_1+\delta, a)$
becomes of order $1$ for $i > 20$. This implies that, in practice, we cannot
calculate with the necessary accuracy $x_{i+1} = F^{(i)}(x_1, a)$ for 
$i > 20$. To address this fundamental limitation, we propose to cut
the sample $s_1, ..., s_N$  into $n_1$
portions of size no more than $n_2=20$, and to treat each portion separately.
This amounts to re-estimating a different initial condition for each such
sub-series, which is a natural step since the sensitivity upon initial
conditions amounts to losing the information on the specific value
of the initial condition.

Our numerical tests show that our MLE works well (see below) by considering
sub-series of size in the range $n_2=4-25$ (for the true value of $a$
equal to the value $1.85$ considered by by McSharry and Smith
\cite{McSharrySmith} that we take as our benchmark for the sake of comparison). 
For larger samples (say, $N=100$), we recommend to cut this sample into
$n_1$ subsamples of size $n_2=4-25$, and treat them separately. It is possible that we lose
some efficiency in treating subsamples separately, but a joint estimation
would require the maximization of the likelihood with the common parameter
$a$ and several different initial value parameters. This procedure would lead
to a very difficult numerical multivariate search problem as any gradient method
would fail due to the very irregular structure of the likelihood function (see below
and figure 1).

The procedure we propose is thus to cut the initial time-series into 
$n_1$ independent subsamples of size $n_2$ in the range $4-25$, and to average the
resulting $n_1$ $a$-estimates. In order to determine 
the optimal value of $n_1$ for a fixed $N$
(say $N=100$) and for the value $a=1.85$ investigated here, we calculate the 
standard deviation sdt$(a)$ over the $n_1$ subsamples as a function of $n_1$.
We find that, basically independently of the noise level $\epsilon$, the pair
$n_1=25, n_2=4$ gives the smallest standard deviation sdt$(a)$.

We have implemented this approach and compared it with the results obtained by 
the method proposed by McSharry and Smith \cite{McSharrySmith}, as discussed
in the next section.

\section{ML version of McSharry and Smith \cite{McSharrySmith} and comparisons}

The main result of McSharry and Smith's paper \cite{McSharrySmith} 
consists in their formulae (13,14) for their proposed
ML cost function. Their idea is to substitute
in the ML cost function
the unknown invariant measure $\mu_a(x)$ of the dynamical system 
(\ref{jjgled}), for a given value of the parameter $a$,
for what should be a realization of the latent
variables $x_i$'s. Notice that $a$ should be
varied in order to determine the maximum likelihood. In practice, the
integral over the unknown invariant measure $\mu_a(x)$ is replaced by a
sum over a model trajectory (which can be calculated since the model is
assumed to be known) of length $\tau \gg N$. Unfortunately, this most
important step is not confirmed by any numerical results (see below).

Before continuing, let us note 
that there is a mistake in the probability density function (pdf)
and likelihood given by their equations (7-9). Using the intuition that
pairs $(s_i, s_{i+1})$ should be used in their equation (5, 6) to
track the deterministic relation between $x_i$ and $x_{i+1} =F(x_i, a)$, we see that
a single latent variable $x_i$ is associated with each pair $(s_i, s_{i+1})$
since $s_i$ is compared with $x_i$ and $s_{i+1}$ with $F(x_i, a)$.
Thus, each $x_i$ is used only once when scanning all possible pairs
$(s_i, s_{i+1})$, for $i=1, ..., N-1$ and in
their ML cost function (13,14). Actually, the correct likelihood
should use only once each {\it observed} random
variable $s_i$, not the latent variable $x_i$. 
Therefore, using pairs $(s_i, s_{i+1})$,  McSharry and Smith
take into account each $s_i, i=2,..., N-1$ twice, and
the end values $s_1, s_N$ once. For $N \gg 2$, their expression (7) is
approximately equals (up to the end terms) to the square of the correct
likelihood. Taking the logarithm in their equation (13) gives approximately twice the
correct likelihood, which gives almost the same estimate as the exact
likelihood. 

While this mistake has no serious consequences for the numerical
accuracy of their calculation for long time 
series $N \gg 2$, it illustrates the difference 
between their construction of the likelihood and our direct approach
presented in the previous section. 
By writing the conditional likelihood for a pair $(s_i, s_{i+1})$ under
a latent variable $x_i$, and by
averaging this conditional likelihood weighted by the invariant measure
$\mu(x|a)$, McSharry and Smith suggest that, by doing so, they
incorporate additional information on the system in question.
If we had a usual probability space, then such averaging would provide
the unconditional likelihood of the pair $(s_i, s_{i+1})$
but, for deterministically chaotic time series, the exact meaning of this averaging is not clear.
Another questionable step of McSharry and Smith is
to multiply these pairwise likelihoods as if the pairs
 $(s_i, s_{i+1})$ were independent. If this was so, this would indeed
 give the unconditional likelihood for the data sample $s_1, ..., s_N$.

But, we deal here with ``deterministic chaos'' which generates not truly random variables
(see for instance \cite{SorArn,Phatak} for discussions 
on the pseudo-randomness nature of such time series). Besides, we have some more information about
the structure of the system in question. Namely, we suppose known
the generating relation (\ref{jjgled}).
This relation contains everything and is, in principle, much more informative than
the stationary invariant measure $\mu(x|a)$ (which is akin to a one-point
statistics while (\ref{jjgled}) contains information on all higher-order
point statistics). Concretely, it is clear that the product
of pdf's for each pair $(s_i, s_{i+1})$ and the resulting likelihood
depends solely on the first initial value $x_1$ since all subsequent
$x_i$ are deterministically determined recurrently.
This remark gives 
the likelihood function (\ref{iwqsm}) in terms of two 
unknown parameters $(a,x_1)$ to be estimated. 
This leads indeed to consider the initial state variable $x_1$ as an unknown parameter to
be estimated (along with $a$) from the sample $s_1, ..., s_N$. 
The likelihood (\ref{iwqsm}) provides a more
detailed form than obtained by averaging over the invariant measure $\mu(x|a)$. 
We can hope that our approach would lead to a more efficient estimate of
$a$. McSharry and Smith avoid the maximization with
respect to $x_1$ in their likelihood (13,14) and replace it by an averaging
over a proxy of the invariant measure. It is doubtful that such a step is
warranted, not speaking of optimality, in view of our numerical tests
presented below.

\begin{figure}[h]
\includegraphics[width=8cm]{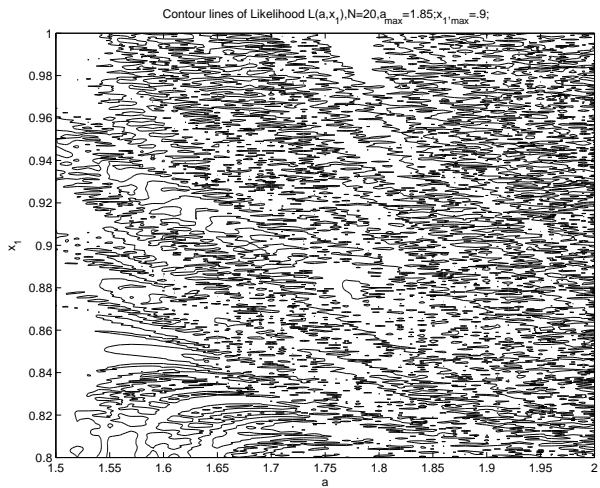}
\caption{\label{Fig1} Contour lines of the ``pure'' log-Likelihood given by expression 
(\ref{iwqsm}) for a given realization of $N=20$ data points generated with a starting value $x_1=0.9$, 
$a=1.85$ and noise std equal to $1$. The log-Likelihood landscape is similar to a 2D
Brownian sheet (2D generalization of a random walk).
}
\end{figure}

We now compare our ``pure'' Maximum Likelihood approach in terms 
of $(a, x_1)$ proposed in section \ref{nmgl} with 
McSharry and Smith's ML method, using numerical tests.
We consider 1000 time series with $N=100$ data points and subdivide 
each of them into
$n_1=25$ sub-series of $n_2=4$ data points. We fix the true $a$
equal to $1.85$ as in \cite{McSharrySmith}  and study 
noises with standard deviations equal to $0.5$ and $1.0$.
Table \ref{table1} shows a significant improvement offered 
by our ``pure'' ML method over McSharry and Smith's average ML, as least for
the set of parameters studied here. 
It is not possible to guarantee that this 
will be the case for all possible parameter values but we believe our method
can not be worse that McSharry and Smith's average ML. A difficulty that should be
mentioned is that the chaotic nature of the dynamics and in particular
the sensitivity of the invariant measure with respect to the control parameter $a$ 
is reflected into an ugly-looking log-Likelihood landscape shown in Figure \ref{Fig1}, with
many competing valleys. Standard numerical methods like gradient or simplex are
unapplicable. We have used a systematic 2D-grid search. Other methods in the field
of computational intelligence, such as stimulated annealing and genetic algorithms, could also
be used.
The sensitivity of the invariant measure with respect to the control parameter $a$
means that the invariant distribution can bifurcate from
an almost uniform distribution on the interval $[-a,1]$ to a
distribution consisting of three delta-functions (this happens around $a \approx 1.75$).

\begin{table}
\begin{center}
\begin{tabular}{|c|c|c|c|c|c|c|c|}
\hline
  noise       &                 & mean$(a)$ & std$(a)$ & $q_1$ & $q_2$ & $q_2-q_1$ & ${\hat \epsilon}$ \\ \hline \hline
std  $0.5$ & Ref.\cite{McSharrySmith}  & $1.816$   & $0.0714$ & $1.630$ & $1.925$ & $0.295$ & \\ \hline
	            & ``pure'' ML    & $1.841$   & $0.0390$ & $1.762$ & $1.913$ & $0.151$  & $0.459$ \\ \hline \hline
std  $1$ &  Ref.\cite{McSharrySmith}    & $1.764$    & $0.123$ & $1.510$ & $1.975$ & $0.465$ & \\ \hline
	            & ``pure'' ML    & $1.885$   & $0.0467$ & $1.781$ & $1.959$ & $0.178$ & $0.766$ \\ \hline \hline
\end{tabular}
\end{center}
\caption{\label{table1} Comparison between McSharry and Smith's ML method \cite{McSharrySmith}
and our ``pure'' ML method described in section \ref{nmgl} over 1000 realizations
of the system (\ref{jjgled}) with true value $a=1.85$ giving 1000 time
series of length $N=100$, each them decorated 
with Gaussian noise with two different standard deviations ($0.5$ and $1$).
$q_1$ and $q_2$ are the sample quantiles at the $2.5\%$ and $97.5\%$ probability level, so that 
$q_2-q_1$ gives the width of the $95\%$ confidence intervals. Our ``pure'' ML method provides us
with an estimation ${\hat \epsilon}$ of the standard deviation of the noise given in the last column.
}
\end{table}

In addition to performing better, our ``pure'' ML approach does not depend on the 
noise level, in contrast with the ML cost function (13,14) proposed by
McSharry and Smith \cite{McSharrySmith}. This is an important advantage when 
the true level of noise is not known (noise error). Our method is insensitive to
such noise error while we have found examples where the optimal estimation of the 
structure parameter $a$ with McSharry and Smith's method
is obtained for a value of the noise standard deviation different from the true value.
In general, the true noise level is not known and McSharry and Smith's method
does not apply in such situation. Our ``pure'' ML method actually provides us with
an estimation ${\hat \epsilon}$ of the standard deviation of the noise given in the last column
of Table \ref{table1}. These estimates have a small bias down (two fitted
parameters were taken into account), which may be due to the fact that 
$n_1$ is not sufficiently large ($n_1=25$; $n_2=4$; $N=n_1 \times n_2=100$).

\section{Discussion of other approaches}

Meyer and Christensen \cite{MeyerChristensen} have proposed to replace the 
ad hoc construction of McSharry and Smith's ML cost function by a Bayesian approach,
assuming noninformative priors for the structural parameter $a$, for the initial
value $x_1$ and for the standard deviation of the noise. Their approach
improves significantly on McSharry and Smith \cite{McSharrySmith} by recognizing
the role of $x_1$ but turns out to be incorrect, as shown by Judd \cite{Judd}, because
their approach amounts to assuming a stochastic model, thus refering to quite another problem.

Based on the formulation of \cite{Berliner}, Judd \cite{Judd} develops a formulation which 
is almost identical to our ``pure'' ML (\ref{iwqsm}) but there are important
distinctions. Similarly to us, Judd introduces $x_1$ but he does not employ it. He
prefers to eliminate the dependency on $x_1$ by averaging this
parameter with a fiducial distribution
(see e.g. \cite{KendallStuart}, Chapter 21, Interval Estimation, Fiducial Intervals).
Judd incorrectly calls the method based on his
equations (4,5) a ML method. In fact, his equations (4,5) gives a
a hybrid of ML, Bayesian and
so-called fiducial methods. It is a ML method with respect to the structural parameter
$a$. It is Bayesian with respect to the initial value $x_1$. It is fiducial since
it does not assume any a-priori density for $x_1$, but uses a prior density
function $\rho(s_1-w)$ (using the notation above) that is in fact a Gaussian density of the noise
with mean value equal to the unknown initial value $s_1$. Using such
density is equivalent to weighting a two-parameter likelihood 
by weights corresponding to different values of noise
disturbances. Thus, the averaged likelihood (5) in \cite{Judd} describes
an ensemble of different noise disturbances of an unknown initial
value $s_1$. This provides a (reasonable but not optimal) method of elimination
of the second parameter $x_1$ from the maximization procedure. It is neither a
pure Bayesian method (that would assume explicitly some a-priori density
for $s_1$ which could be arbitrary, and not necessarily equal to
$\rho(s_1-w)$), nor a ML method for two unknown parameters as we suggested above in
section \ref{nmgl}.

In this context in view of the emphasis on Bayesian methods to solve this problem
\cite{MeyerChristensen,Judd}, it is perhaps useful to stress that 
the probability theory rule $P\{A,B\}=P\{A|B\}~ P\{B\}$
is often freely called ``the Bayes rule.'' This is why the averaging
of likelihoods over conditional state variables can be called 
Bayesian approaches, although this is not quite correct
since the latent (state) variables are not random values
in the standard meaning of this notion (as it is assumed by 
McSharry and Smith), although the state variables have a limit 
invariant measure, as we said above. The Bayesian
approach assumes that parameters are random values. For instance, 
McSharry and Smith assume that the latent (state) variables
are random variables, which is not quite so,
although the state variables have a limit invariant measure, as we said above.
We stressed already that the series of state variables can be considered as 
a degenerate set of random values that are determined by one single random variable,
namely $x_1$. What is more natural? To consider $x_1$ as a random variable with
a distribution determined by the invariant measure, or to consider $x_1$ as
an unknown parameter to be estimated? The answer, in our opinion, is dictated
by consideration of efficiency: the different examples that we have explored
suggest that the latter is as a rule more efficient (has smaller mean square error), 
at least for some combinations of sample size $N$ and noise level.

As all the above has shown, the major obstacle is the loss of information 
on the initial value $x_1$ by the unstable logistic map beyond $10-25$ time steps.
We proposed the simple recipe of cutting the time series in short pieces and
of averaging the estimations. Judd proposes a shadowing method \cite{Judd}. It is not obvious
that this will result in a consistent estimation and that this will 
overcome the intrinsic difficulty in
treating long realizations (which is a necessary condition for unbiased estimations). 

In sum, there is no analytical proof of consistency for all the
estimation methods discussed until now (including the suggestions performed by the most
convincing work to date \cite{Judd} and our ``pure'' ML). It is useful to 
analyze the only method to our knowledge for which one can derive a proof of consistency
in the present context, that is, the method of statistical moments.

\section{The method of statistical moments}

The method of statistical moments provides a
consistent estimate of the parameters for non-linear maps with ergodic
properties. The method of statistical moments is the unique theoretically
proven consistent estimator among all methods suggested so far by
other authors. Although the moment estimates are known to have little
efficiency, they are consistent! Consistency of all estimates suggested
earlier including ours above were confirmed only numerically, which is very
dangerous for instable non-linear maps. 

We consider four moment of the observed time series: 
$\langle s \rangle_N, \langle s^2 \rangle_N, \langle s^3 \rangle_N$
and $\langle s_i s_{i+1} \rangle_N$, where the brackets stand for time
averaging over some time interval $N$.
Building on the knowledge that the series $\{x_i\}$ is ergodic \cite{Collet} and 
using (\ref{mgmwlw},\ref{jjgled}), we obtain the following relations
\ba
\langle s \rangle_N  &\to& \langle x \rangle_{\infty}~, \label{1} \\
\langle s^2 \rangle_N  &\to& \langle x^2 \rangle_{\infty}~,  \label{2}\\
\langle s^3 \rangle_N  &\to& \langle x^3 \rangle_{\infty} + 3 \langle x \rangle_{\infty} \epsilon^2~, 
\label{3} \\
\langle s_i s_{i+1} \rangle_N &\to& \langle x \rangle_{\infty} - a \langle x^3 \rangle_{\infty}~.
\label{4}
\ea
Besides, averaging equation (\ref{jjgled}), we get
\be
\langle x \rangle_{\infty} = 1 - a \langle x^2 \rangle_{\infty}~.  \label{5}
\ee
This provides us with five limit relations (\ref{1}-\ref{5})
with five unknown parameters: $a, \langle x \rangle_{\infty}, \langle x^2 \rangle_{\infty},
\langle x^3 \rangle_{\infty}$ and $\epsilon$. Solving these five relations with respect to
the unknown parameters, we get the so-called estimates of the
method of moments:
\ba
{\hat a} &=& {\langle s_i s_{i+1} \rangle_N + 2 \langle s \rangle_N - 3(\langle s \rangle_N)^2
\over 3 \langle s \rangle_N \langle s^2 \rangle_N - \langle s^3 \rangle_N}~, \label{6}\\
\langle {\hat x} \rangle_{\infty} &=& \langle s \rangle_N ~, \label{7}\\
\langle {\hat x}^2 \rangle_{\infty} &=& \langle s^2 \rangle_N - {\hat \epsilon}^2~, \label{8}\\
\langle {\hat x}^3 \rangle_{\infty} &=& {1 \over {\hat a}} \left(\langle s \rangle_N - 
\langle s_i s_{i+1} \rangle_N \right) ~,  \label{9}\\
{\hat \epsilon}^2 &=& {\langle s^3 \rangle_N - \langle x^3 \rangle_{\infty} 
\over 3 \langle s \rangle_N}~.  \label{10}
\ea
Because of the limit relations (\ref{1}-\ref{4}) (which are valid because of
the ergodicity of the time series $\{x_i\}$ \cite{Collet}), the estimates 
(\ref{6}-\ref{10}) are consistent if $N \to \infty$.

\begin{table}
\begin{center}
\begin{tabular}{|c|c|c|c|c|c|c|}
\hline
sample size & Noise std  & Estimate          & $q_1$   & $q_2$   & $q_2-q_1$ \\ 
$N$         & $\epsilon$ & $(a) \pm$ std     &         &         & 		     \\ \hline \hline
$100$		& $0.05$     & $1.8768\pm 0.0926$& $1.684$ & $2.000$ & $0.316$   \\ \hline
$1000$		& $0.05$     & $1.8544\pm 0.0418$& $ 1.774$ & $1.936$ & $ 0.162$   \\ \hline
$10000$		& $0.05$     & $1.8503\pm 0.0136$& $ 1.824$ & $1.878$ & $ 0.054$   \\ \hline
$100000$	& $0.05$     & $1.8499\pm 0.0044$& $ 1.842$ & $1.858$ & $ 0.016$   \\ \hline \hline
$100$		& $0.1$     & $1.8456\pm 0.1546$ & $1.499$ & $2.000$ & $0.501$   \\ \hline
$1000$		& $0.1$     & $  1.8532 \pm 0.0815$ & $1.693$ & $2.000$ & $0.307$   \\ \hline
$10000$		& $0.1$     & $1.8505\pm 0.0279$ & $1.795$ & $1.908$ & $0.113$   \\ \hline
$100000$	& $0.1$     & $1.8497\pm 0.0089$ & $1.833$ & $1.867$ & $0.034$   \\ \hline \hline
$100$		& $0.5$     & $1.2411\pm 0.7331$ & $0$ & $2.000$ & $2.000$   \\ \hline
$1000$		& $0.5$     & $1.6907\pm 0.3496$ & $ 0.903$ & $2.000$ & $1.097$   \\ \hline
$10000$		& $0.5$     & $1.8244\pm 0.1659$ & $ 1.467$ & $2.000$ & $0.533$   \\ \hline
$100000$	& $0.5$     & $1.8554\pm 0.0741$ & $1.715$ & $2.000$ & $0.285$   \\ \hline
\end{tabular}
\end{center}
\caption{\label{table2} Estimation of the structural parameter $a$ by 
the method of statistical moments (expression (\ref{6})) for the logistic map 
$x_{i+1} = 1 - a x_i^2, a = 1.85$;  the observations are $s_i = x_i + \eta_i; \eta_i$
is a Gaussian random variable $N(0,\epsilon)$. As in table \ref{table1},
$q_1$ and $q_2$ are the sample quantiles at the $2.5\%$ and $97.5\%$ probability level, so that 
$q_2-q_1$ gives the width of the $95\%$ confidence intervals.
Each estimate for $a$ and std are based on 1000 simulated samples. }
\end{table}

We present in table \ref{table2} the estimates of the parameter $a$ given by expression 
(\ref{6}). The consistency of the method of statistical moments is clearly suggested 
by the numerical results, as seen
from the bracketing of the true value by $(a) \pm$ std and by $q_1$ and $q_2$.
However, as we already pointed out, the method of statistical moments is rather inefficient:
the ratio of its standard deviation for $a$ to that of the ``pure'' ML is about $4$ for
$N=100$ and $\epsilon =0.1$ for instance.

\section{Concluding remarks}

We have proposed a ``pure'' Maximum Likelihood  (ML) method to estimate the
structural parameter of a deterministically chaotic low-dimensional
system (the logistic map), which adds the initial value $x_1$ to the structural
parameter to be determined. We have compared quantitatively this method with the ML method 
proposed by McSharry and Smith \cite{McSharrySmith} based on an averaging
over the unknown invariant measure of the dynamical system. 
A key aspect of the implementation of our approach lies in the compromise
between the need to use a large number of data points for the ML to become
consistent and the unstable nature of dynamical trajectories which loses exponentially
fast the memory of the initial condition. This second aspect prevents using our 
``pure'' ML for systems larger than $10-25$ data points. For larger time series, we
have found convenient to devide them into subsystems of very small lengths and then
to average over their estimations. Numerical tests suggest that this direct ML method
provides often significantly better estimates than previously proposed approaches.

The difference between McSharry and Smith's averaging over the invariant measure
and our ``pure'' ML is reminiscent of the distinction between ``annealed'' versus
``quenched'' averaging in the statistical physics of random systems, such as spin glasses
\cite{Mezard,dsbook}. It has indeed been shown that the correct theory of 
strongly heterogeneous media is obtained by performing the thermal Gibbs-Boltzmann
averaging over fixed structural disorder realizations, similarly to our use of a specific
trajectory of the latent variables $x_i$'s. In constrast, performing the 
thermal Gibbs-Boltzmann averaging together with an averaging over different realization
of the structural disorder describes another type of physics, which is not that
of fixed heterogeneity. This second incorrect type of averaging is similar to the
averaging of the ML over the invariant measure performed by McSharry and Smith.

There are several ways to improve our approach. One simple 
implementation is to use overlapping running windows. Another method is to 
re-estimate the realized trajectory by using the extended Kalman filter method
(however, difficulties may arise due to the existence of a maximum in the logistic map).
Using shadowing methods as proposed in \cite{Judd} in our context would also
be interesting to investigate.

Let us end with a cautionary note.
As we just said, the ML approach for two parameters $(a,x_1)$ that we suggest here
evidently works only for a limited sample size N (perhaps, $N<25$ or
so) due to the sensitivity upon initial conditions of the 
chaotic logistic map. As is well-known in classical statistics, 
ML-estimates have a bias that
can be considerable if $N$ is not large (say, $N<100$ or so). The
ML-estimates are usually only asymptotically unbiased. Thus, for $N=25$ (and
all the more for $N=4$), ML-estimates can exhibit a considerable bias. Thus,
averaging biased estimates as we proposed many not result in a consistent estimation.
Therefore, we cannot assert that our ML method (as well as any other suggested
methods) is consistent. We can only observe, for particular
combinations of the considered parameters, the numerically determined mean square error of
our suggested estimates with respect to the true parameter value.
We are pleased if these errors are not too high, although our
estimates can be biased (though, with small bias). But we are not able
to make such bias arbitrarily small by increasing the sample size $N$, due to
the instability under the iterations of the logistic map which leads to a loss
of information about the initial value $x_1$.
Thus, the situation is rather hopeless for the establishment of 
a meaningful statistical theory of estimation using the continuous
theory of classical statistics to such discontinuous objects as the invariant measures
of chaotic dynamical systems.

\begin{acknowledgments}
We are grateful to K. Ide for useful discussions.
This work is partially supported by a LDRD-Los Alamos grant and by
the James S. Mc Donnell Foundation 21st century scientist
award/studying complex system. 
\end{acknowledgments}

\section*{Appendix A: One-step and total least-square estimations}

McSharry and Smith noticed that the one-step leasts-square method
gives strongly biased results for the estimation of $a$ \cite{McSharrySmith}. Indeed, 
the method of estimation of the parameter $a$ by the
one-step least square method is evidently inconsistent, since the
deviations (of the random variables) to be minimized in a least-square sense are
\ba
s_{i+1} - F(s_i,a) &=& x_{i+1} + \eta_{i+1} - F(x_i +\eta_i ,a)  \nonumber   \\
&=& \eta_{i+1} + 2a x_i \eta_i +a \eta_i^2~,
\ea
which has non-zero expectation equal to $a \epsilon^2$. 
But, the fundamental least-square
principle consists in the minimization of deviations with zero mean. There
are no least-square schemes that would suggest to minimize random
deviations with non-zero mean depending on an unknown parameter. Thus, it is
not reasonable to include the least-square method in any reasonable comparison.

The method called by McSharry and Smith as ``total least-squares'' (TLS) is
applied in situation when the variables $x_i$ are known only with some errors 
$\eta_i$. This situation is called in statistics a ``Confluence analysis,''
or ``Estimation of a structural relation between two (or more) variables
in the presence of errors on both variables'' \cite{FrischR,Geary,KendallStuart}.
In such a situation of confluence analysis, since the $x_i$'s
are in fact unknown (nuisance) parameters whose
number grows with sample size, there is no guarantee of consistency of
the ML estimates of the structural parameter $a$. 

As an example, let us consider the very simple confluent scheme:
\ba
Y_i &=& X_i + \eta_i~,    \\
Z_i &=& X_i + \zeta_i ~.
\ea
Suppose we observe a sample of $N$ pairs $(Y_i , Z_i), i=1, ..., N$, where $X_i$ are
unknown arbitrary values and $\eta_i, \zeta_i$  are iid Gaussian random variables with 
standard deviation $\epsilon$.
The problem consists in estimating the parameter $\epsilon$. 
Similarly to the situation with (\ref{mgmwlw}) and (\ref{jjgled}) studied in 
\cite{McSharrySmith}, no restrictions are
placed on the $X_i$'s. The likelihood $L(\epsilon,X_1, ..., X_N | (Y_i,Z_i), i=1,...,N)$ is
$$
L(\epsilon,X_1, ..., X_N | (Y_i,Z_i), i=1,...,N) \propto
$$
\be
\epsilon^{-2N}~ \exp \left[ -(1/2\epsilon^2) \sum_{i=1}^N (Y_i - X_i)^2  
-(1/2\epsilon^2) \sum_{i=1}^N (Z_i - X_i)^2 \right] ~.
\label{nhhjw}
\ee
The MLE ${\hat X}_i$'s of the $X_i$'s  (that coincide 
in this case with the least-square estimates) are:
\be
{\hat X}_i =  {Y_i + Z_i \over 2}~.
\label{mgfmlr}
\ee
Inserting (\ref{mgfmlr}) into (\ref{nhhjw}), we get
\ba
{\hat L}(\epsilon | (Y_i,Z_i), && i=1,...,N) \propto  \nonumber \\
&& \epsilon^{-2N}~ \exp \left[ -(1/4\epsilon^2) \sum_{i=1}^N (Y_i - Z_i)^2 \right] ~.
\label{nhsssjw}
\ea
Thus, the MLE of the parameter $\epsilon$ obtained from (\ref{nhsssjw}) satisfies 
\be
\epsilon^2 = {1 \over 4N}  \sum_{i=1}^N (Y_i - Z_i)^2 ~.
\label{njgtje}
\ee
Since ${\rm E}\left[ (Y_i - Z_i)^2 \right] = 2 \epsilon^2$, the estimate 
(\ref{njgtje}) is inconsistent. A
consistent (``corrected'') estimate is
\be
\epsilon^2 = {1 \over 2N}  \sum_{i=1}^N (Y_i - Z_i)^2 ~.
\label{njgtbbbje}
\ee
Thus, we see that the MLE of the structural parameter $\epsilon$ is inconsistent due
to the increasing number of nuisance parameters. Thus, the direct use of the least-square (or
total least-square) in the confluent situation is not justified, and was not recommended in
any statistical textbook. Instead, standard
statistical works recommend a ``corrected''
ML estimates (see for instance \cite{Geary,KendallStuart}).

We should stress in addition that there is a significant difference between
the standard confluent analysis and the problem addressed in \cite{McSharrySmith}.
Confluent analysis
deals with arbitrary unknown (distorted)
arguments $x_i$, whereas in \cite{McSharrySmith}, the latent variables
$x_i$ are related by the non-linear map (\ref{jjgled}). 
The information on the
structure of the $x_i$'s is not used in Confluence Analysis while it can really
help in the estimation procedure as shown in \cite{McSharrySmith} and in the present work.


\vskip -0.6cm
\begin{thebibliography}{}

\bibitem{McSharrySmith} P.E. Mcsharry and L.A. Smith, Phys. Rev. Lett. 83, 4285 (1999)

\bibitem{FrischR}  Frisch R. Statistical Confluence Analysis by Means of Complete
 Regression Systems, Oslo, 1934.

\bibitem{Geary} Geary R.C. Non-linear functional relationship 
between two variables when one variable is controlled, 
J. Amer. Statist. Ass. 48, 94 (1953).

\bibitem{KendallStuart} M. Kendall and A. Stuart, 
The advanced theory of statistics,
Curvilinear Dependencies, 2d ed. (New York, Hafner Publ. Co., 1961),
Chapter 29, Section 29.50.

\bibitem{SorArn}  D. Sornette and A. Arn\'eodo, J. Phys. (Paris) 45, 1843 (1984).

\bibitem{Phatak} S.C. Phatak and S.S. Rao,
Phys. Rev. A. 51 (4 Part B), 3670 (1995). 

\bibitem{MeyerChristensen} R. Meyer and N. Christensen,
Phys. Rev. E 62, 3535 (2000).

\bibitem{Judd} CK. Judd, Phys. Rev. E 67 (2), 026212 (2003).

\bibitem{Berliner} M.L. Berliner, J. Am. Stat. Assoc. 86, 939 (1991).

\bibitem{Collet} P. Collet and J.-P. Eckmann, 
Iterated maps on the interval as dynamical systems 
(Basel; Boston: Birkhauser, 1980).

\bibitem{Mezard} M., M\'ezard,~M.,~Parisi,~G. and Virasoro,~M.,  Spin Glass Theory
and Beyond (World Scientific,~Singapore, 1987).

\bibitem{dsbook} D. Sornette,
Critical Phenomena in Natural Sciences
(Springer Series in Synergetics, Heidelberg, 2000), see chapter 16.

\end{thebibliography}
\end{document}